\documentclass[12pt]{article}
\usepackage{epsfig}
\usepackage{rotating}

\def\gray{\special{ps: 0.4 setgray}}
\def\black{\special{ps: 0.0 setgray}}

\newcommand{\draft}{
\newcount\timecount
\newcount\hours \newcount\minutes  \newcount\temp \newcount\pmhours
 
\hours = \time
\divide\hours by 60
\temp = \hours
\multiply\temp by 60
\minutes = \time
\advance\minutes by -\temp
\def\hour{\the\hours}
\def\minute{\ifnum\minutes<10 0\the\minutes
            \else\the\minutes\fi}
\def\clock{
\ifnum\hours=0 12:\minute\ AM
\else\ifnum\hours<12 \hour:\minute\ AM
      \else\ifnum\hours=12 12:\minute\ PM
            \else\ifnum\hours>12
                 \pmhours=\hours
                 \advance\pmhours by -12
                 \the\pmhours:\minute\ PM
                 \fi
            \fi
      \fi
\fi
}
\def\fullclock{\hour:\minute}
\begin{centering}
\gray
\begin{sideways}
\font\Hugett  =cmtt12 scaled\magstep2
{\Hugett Draft: \today,\clock}
\end{sideways}
\black
\end{centering}
\vskip -1.7cm
$\phantom{a}$
} 
\catcode`\@=11 
\def\lsim{\mathrel{\mathpalette\@versim<}}
\def\gsim{\mathrel{\mathpalette\@versim>}}
\def\@versim#1#2{\vcenter{\offinterlineskip
        \ialign{$\m@th#1\hfil##\hfil$\crcr#2\crcr\sim\crcr } }}
\catcode`\@=12 

\def\nextline{\hfill\break}

\def\mycomm#1{\nextline\strut\kern-3em{\tt ====> #1}\nextline}

\def\nextline{\hfill\break}
\def\tstrut{\vrule height 1.2em depth 0.5em width 0pt}
\newcommand{\beq}{\begin{equation}}
\newcommand{\eeq}{\end{equation}}
\def\eqref#1{(\ref{#1})}

\def\barq{\bar q}

\def\bars{\bar s}

\def\Mred{\hbox{$\mu_{\scriptscriptstyle{\rm red}}$}}

\begin{document}
\vskip1.5cm
\begin{center}
{\Large\bf The doubly heavies:\\
\tstrut
$\bar Q Q \bar q q$ and
$Q Q \bar q \bar q$ tetraquarks and $QQq$ baryons}
\end{center}
\medskip
\begin{center}
{\bf Marek Karliner$^1$}\footnote{\tt marek@proton.tau.ac.il}
\\
and 
\\
{\bf Shmuel Nussinov$^{1,2}$}\footnote[7]{\tt nussinov@post.tau.ac.il}
\end{center}
\vskip1cm
\begin{center}
$^1$Raymond and Beverly Sackler School of Physics and Astronomy\\
Tel Aviv University, Tel Aviv, Israel
\\
\strut\\
$^2$Institute for Quantum Studies \\
Chapman University, Orange, CA 92866, USA
\end{center}
\begin{abstract}
\noindent
Recent discoveries by Belle and BESIII of charged exotic 
quarkonium-like resonances provide fresh impetus for study of heavy exotic 
hadrons.  In the limit $N_c\to\infty$, $M_Q \to \infty$ the $\bar Q Q 
\bar q q'$ tetraquarks \hbox{(TQ-s)} are expected to be narrow and slightly below or 
above the $\bar Q q'$ and $Q \bar q$ two-meson threshold.  The isoscalar 
TQ-s manifest themselves by decay to $(\bar Q Q)\,\pi\pi$, and the 
${\sim}30$ MeV heavier charged isotriplet TQ-s by decays into $(\bar Q Q)\pi$.
The new data strongly suggest that the real world with $N_c=3$, $Q=c,b$ 
and $q,q' = u,d$ is qualitatively described by the above limit. 
We discuss the relevant theoretical estimates and suggest new 
signatures for TQ-s in light of the recent discoveries. 
We also consider ``baryon-like" states $Q Q' \bar q\bar q'$, 
which if found will be direct evidence not just for near-threshold binding of 
two heavy mesons, but for genuine tetraquarks with novel color networks.
We stress the importance of experimental search for doubly-heavy baryons
in this context.
\end{abstract} 
\vskip-10cm 
\vfill 
\eject   

It has been realized early on that quark models and QCD  sustain
a much richer pattern of different multi-quark and/or color network
configurations, beyond the ``non-exotic" standard $\bar q q$ mesons and 
$qqq$ baryons. Still, production rates of such particles 
are often suppressed and the  light pions 
will in most cases allow rapid decays of the exotics 
into final states with pion(s) -- turning them into very broad 
resonances. This explains why the vast majority of known hadrons
\cite{PDG} are just simple mesons and baryons. There is, however,
growing evidence for the existence of exotic mesons containing both
heavy and light quark-antiquark pairs -- $\bar c c \bar q q$ \cite{X3872}
and as well as manifestly exotic $\bar b b \bar d u$ \cite{Karliner:2008rc}, 
\cite{Belle:2011aa} and most recently
$\bar c c \bar d u$ \cite{Ablikim:2013zna}.

In a recent  article \cite{Weinberg:2013cfa} S.~Weinberg made prescient
observations on $qq \bar q \barq $  tetraquarks  (TQ-s) in the
large-$N_c$ limit, correcting some statements in S.~Coleman's  Erice
Lectures $\cite{ColemanErice}$. Specifically, TQ-s can be created from
the vacuum,  as in lattice calculations \cite{Alford:2000mm}, in two
ways, via products of color-singlet $\bar q q$ bilinears: 
\beq | {\cal T} \rangle = [\bar q_1(x) \Gamma_I  q_2(x)] \cdot
[\barq_3(x) \Gamma_J q_4(x)] | 0 \rangle \label{bilinearA} \eeq   
and
the ``crossed" $2\Leftrightarrow 4$ version, with $\Gamma$ indicating
a Dirac algebra matrix.  TQ-s with a specific value of isospin can be
constructed by taking linear combinations of the $\{1,2\}$ or $\{3,4\}$
terms with their $\{2,1\}$ or $\{4,3\}$   permutation, respectively.

The rapid two-pion decay of $f_0(980)$ and $\eta \pi$ decay of $a_0
(980)$ are avoided if, (as first suggested in a bag model context by
Jaffe \cite{Jaffe:1976ih}), the latter are   $\bars s  (\bar u u \pm
\bar d d)$ TQ-s. The above decays are then slowed by Zweig's rule, 
$1/{N^2 }$ suppression of $\bar s s\rightarrow \bar u u$/$\bar d d$ transitions, 
and the ``natural" decay channel  $\bar K K $  is above 980 MeV.

These features stem from the fact that the strange quark 
is somewhat heavier than the light $u$ and $d$ quarks,
making the $f_0(980)$ and $a_0(980)$ the harbingers of 
recent and likely future TQ-s, 
$\bar Q Q \bar q q $  with $Q = c$,$b$;  $q=u$,$d$.
Such systems have an additional relevant small parameter, 
namely $m_q/M_Q \sim 1/5$ and $\sim 1/15$, respectively.
Rearrangement of $\bar Q Q  \bar q q$ subsystems with 
subsequent decay into $J/\psi$ (or $\Upsilon$) and pion(s)  
is always allowed by kinematics, even when the TQ is
lighter than the sum of the masses of the $\bar Q q + Q\bar q$ mesons, 
but such rearrangements are $1/N_c$ suppressed. 

This suppression is the key to the relatively small width of the 
manifestly exotic charged resonances recently discovered
 in both $\bar b b$ by Belle \cite {Belle:2011aa} -- and just
 a few days ago, the remarkable peak $Z_c(3900)$ 
at $3899.0\pm3.6\pm4.9$ MeV with $\Gamma=46\pm10\pm20$  MeV
in $\bar c c$ system by BESIII
\cite{Ablikim:2013zna}.\footnote{Note that a ${\cal O}(30)$
MeV energy implies a velocity of ${\sim}(1/8)$c  
for  $\bar D$ and $D^*$ in the final state.
For a system of size 
$\sqrt{m_D\times 30~{\rm MeV}}\sim{\cal O}(0.8)$ fm, 
this implies traversal time of
6.4 fm/$c$,
corresponding to a crude estimate of the $S$-wave state width
$\Gamma\simeq c/(6.4\,{\rm fm}) = 200~{\rm MeV}/6.4
\sim 30~{\rm MeV}.$
The fact that  this is close to the
observed width suggests that we indeed have a loosely-bound $\bar D D^*$ 
resonance decaying mainly into its constituent mesons.}

The relatively slow decay of these exotic resonances implies that the
dominant configuration of the $\bar Q Q \bar q q$  four-body system is
{\em not} that of a low-lying $\bar Q Q$ quarkonium and pion(s).  
The latter have a much lower energy than the respective 
two-meson thresholds $\bar M M^*$ and $\bar{M^*} M^*$, $(M=D,B$),
but do readily fall apart into $(\bar Q Q)$ and pion(s) and would result in very large 
decay widths.

Instead, just like in the case of $f_0 (980)$ and $a_0 (980)$, we should
 view these systems as loosely bound states and/or near threshold resonances in the two heavy-meson system.

Such ``molecular" states, $\bar D D^*$, etc., were introduced
in Ref.  \cite{Voloshin:1976ap}. They were  later extensively discussed 
\cite{deusons,Tornqvist:2004qy} 
in analogy with the deuteron which binds via exchange of pions and
other light mesons, and were referred to as ``{\em deusons}". 
The key observation is that the coupling to the heavy mesons of the light mesons  exchanged 
($\pi$, $\rho$, etc.) becomes universal and independent of $M_Q$ for $M_Q\to\infty$,
and so does the resulting potential in any given $J^{CP}$ and isospin channel.
In this limit the kinetic energy $\sim p^2/(M_Q)$ vanishes, 
and the two heavy mesons bind with a  binding energy $\sim$ the maximal depth of the attractive meson-exchange potential.  

For a long time it was an important open question whether these 
consideration apply in the real world with large but finite masses of the 
$D$ and $B$ mesons.
The recent experimental results of Belle \cite {Belle:2011aa} 
and  BESIII \cite{Ablikim:2013zna}, together with theoretical analysis 
in Refs.~\cite{Karliner:2011yb} and \cite{Karliner:2012pc} 
strongly indicate that 
such exotic states do exist -- some were already found and more 
are predicted below.

Due to parity conservation, the pion cannot be  exchanged  in the $\bar M M$
system,  but it does contribute in the $\bar M M^*$ and $\bar{M^*} M^*$ channels. 
The \hbox{$\vec \tau_1\cdot\vec\tau_s$} isospin nature of the exchange  implies that 
the the binding is 3 times stronger in the isoscalar channel,
experimentally accessed in the decay 
into $\Upsilon\pi\pi$ or $J/\psi\pi\pi$, than in the
isovector channel, seen in decay into $\Upsilon\pi$ or $J/\psi\pi$.
It was estimated \cite{Karliner:2011yb,Karliner:2012pc}
that in the bottomonium system 
this difference in the binding potentials raises the $I{=}1$ exotics by up to
$40$-$50$ MeV above the $I{=}0$ exotics. In the charmonium system this splitting 
is expected to be slightly smaller, because the $\bar D D^*$/$\bar{D^*} D^*$ 
states are larger than $\bar B B^*$/$\bar{B^*} B^*$.
This is because
the reduced mass in the $\bar B B^*$ system is approximately 2.5 times
larger than in the $\bar D D^*$ system. On the other hand, the net attractive
potential due to the light mesons exchanged between the heavy-light mesons
is approximately the same, since $m_c, m_b \gg \Lambda_{QCD}$.
As usual in quantum mechanics, for a given potential 
the radius of a bound state or a resonance gets
smaller when the reduced mass grows, so the 
$\bar D D^*$ states are larger than the $\bar B B^*$ states.
Because of this difference in size
the attraction
in both $I{=}0$ and $I{=}1$ charmonium 
channels is expected to be somewhat smaller.

We are interested in the dependence of the binding energy on the heavy meson masses 
in the  $\bar M M$, $\bar M M^*$ and $\bar{M^*} M^*$  systems controlled 
by Hamiltonians of the form:
\beq
H = a\cdot p^2 + V(r)
\label{hamiltonian}
\eeq
\noindent
where $a\equiv 1/2 \Mred$, with \Mred\ the reduced mass and the channel-specific
potential $V(r)$ is assumed to be independent of the heavy-meson mass. 

For Hamiltonians linear in a parameter $a$, the ground-state energy $E_0(a)$ is 
a convex function of $a$, since second-order perturbation makes a negative contribution
to the ground state. 
For  $H= \sum_i^n a_i H_i$, the function
$E_0( a_1,a_2, \ldots, a_n)$ describes a convex surface in $n$ dimensions. 
With additional theoretical inputs or experimental results, 
the $n$-dimensional convexity might help estimating how the binding 
is affected by scaling various pieces of the potential.
 
Convexity implies that  $E_0(a)$ at any point in the  $(0,A)$  interval  
exceeds the linear interpolation
\beq
 E_0(a) > E_0(0) + a \cdot {E_0(A) - E_0(0) \over A}
\label{LinInterp}
\eeq
Related arguments were used to prove flavor-dependent baryon-meson 
mass inequalities \cite{Nus1,Lieb}. Convexity was also used to predict the masses
of doubly-heavy $b\bar c$ mesons \cite{Nussinov:1979di}.

Convexity  is not affected by linear transformations of the $a_i$ and we can use 
any unit of inverse mass to convert the parameter $a \sim 1/M_Q$ into 
a dimensionless parameter $\tilde a$. 

Since we are comparing with the pion and with other light
exchanges, we use the ``natural" unit of $\sim 0.8$ Fermi  $\sim 4.0\ {\rm GeV}^{-1}$
for the range and width of the corresponding potentials.

With  $m_D \sim 2$ GeV and $m_B \sim 5.3$ GeV the relevant dimensionless 
parameters $\tilde a(D) = 1/8$ and $\tilde a(B) = 1/21$ are indeed small, 
suggesting that even the first-order perturbation linear interpolation 
may be satisfactory.

In the infinite mass case, corresponding to $\tilde a=0$, the relative distance between
the mesons is effectively fixed at the minimum of the potential and the binding 
$E_0(0)$ is  then just the attractive $V_{min}$  at this position.

We can use the existing data in order to make a very rough estimate of
the isovector binding potential which is  $E_b^{I=1}$ in the $m_Q \to \infty$ limit, i.e. at
$\tilde a=0$.
We have two data points: $Z_c(3900)$ at
$\tilde a(D)$ is approximately 27 MeV above $\bar D D^*$ threshold and
$Z_b(10610)$ at $\tilde a(B)$ 
is approximately 3 MeV above $\bar B B^*$ threshold. Linear
extrapolation to $\tilde a =0$ yields $E_b^{I=1}(\tilde a{=}0)\approx {-}11.7$ MeV. In
view of the convexity, the actual binding energy is likely to slightly exceed this linear extrapolation.

We can then use this result for the isovector channel to estimate the $\bar B
B^*$ binding in the isoscalar channel. Assuming that the isoscalar binding energy
in the $m_Q\to\infty$ limit is 3 times larger than for the isovector,
i.e. $E_b^{I{=}0}(\tilde a{=}0) \approx 3\cdot({-}11.7)={-}35$ MeV. 
\,$X(3872)$ is at $\bar D D^*$ threshold,
providing an additional data point of $E_b^{I{=}0}(\tilde a(D))\approx 0$ 
in the isoscalar
channel.\footnote{For simplicity we assume $X(3872)$ is an isoscalar, since it
has no charged partners, and we ignore here the issue of isospin breaking in
its decays.}
Linear extrapolation to $\tilde a(B)$ yields approximately ${-}20$
MeV as the $\bar B B^*$ binding energy in the isoscalar channel.

The upshot is that the newly discovered $Z_c(3900)$ isovector resonance
confirms and refines the estimates in
\cite{Karliner:2011yb,Karliner:2012pc} for the mass of
the putative $\bar B B^*$ isoscalar bound state. 
This immediately leads to several predictions:
\begin{itemize}
\item[a)] two $I=0$ narrow resonances in the bottomonium system,
about 23 MeV below $Z_b(106010)$ and $Z_b(10650)$, i.e. about 20 MeV below 
the corresponding $\bar B B^*$ and $\bar{B^*} B^*$ thresholds;
\item[b)] an $I=1$ resonance about 30 MeV above $\bar{D^*} D^*$ threshold;
\item[c)] an $I=0$ resonance  near $\bar{D^*} D^*$ threshold.
\end{itemize}

Various authors, e.g.~\cite{Ader:1981db},\cite{Manohar:1992nd},\cite{Gelman} 
considered  $QQ \bar q_1 \bar q_2$ \,TQ-s.
If such states do exist, producing and discovering even the
lightest $cc\bar u \bar d$ is an extraordinary challenge.  One needs to
produce  {\em two} $\bar Q Q$  pairs 
and then rearrange them, so as to form
$QQ$ and $\bar Q \bar Q$ diquarks, rather than the more favorable
configuration of two $\bar QQ$ and color singlets.  
{\em Then}  the $QQ$ diquark needs to pick up a
$\bar u\bar d$ light diquark, rather than a $q$, to make a $QQq$ baryon,
suppressing the production rate of these TQ-s below the rate  of $QQq$
production.

A small ray of hope comes from the observation of the doubly-heavy
$B_c = (\bar b c)$ mesons \cite{Bc}, suggesting
that  simultaneous production of $\bar b b$ and $\bar c c$ pairs 
which are close to each other in space and
in rapidity and can coalesce to form doubly-heavy hadrons is not too
rare. For example, in the last paper in Ref.~\cite{Bc} CDF reported
$108\pm15$ candidate events in the $B_c^{\pm} \to J/\psi \pi^\pm$
channel.
This is an encouraging sign for the prospects of producing and observing the
$ccq$ and $bcs$ baryons and hopefully also the $cc\bar u\bar d$ \,TQ. 
ATLAS and CMS and especially LHCb probably have the best chance 
of discovering these states.

If the new TQ lies below say, the $D D^*$ threshold, it will be stable under the 
strong interaction and will decay only weakly.\footnote{A statement slightly 
modified if the TQ mass is above the 
$2 m_D$ threshold, as it can then first decay through an exceedingly
narrow EM decay $D^*\to D^0 \gamma$
\cite{PDG},\cite{Nussinov:1998se},
followed by the weak decays of the the $D$-s \cite{Gelman}.}
As we discuss below, if they lay above threshold, 
they may still may manifest themselves as narrow $D D^*$ resonances.

The following theoretical considerations are relevant for the last
point above.
First, in clear contradistinction to the states discussed earlier, the new
TQ-s are {\em unlikely} to be molecular. This is because both heavy mesons
contain light antiquarks, rather than a $\bar q$ in one and a $q$ in the other,
causing the $\omega$ and $\rho$ exchange to be repulsive, 
rather than attractive.
With no bound  $\bar D D^*$ states, one might expect the
new $D D^*$ states to lay higher and manifest themselves 
as broad  resonances, at best. 
It is interesting to point out that the above conclusion is at 
odds with results of a coupled-channel analysis carried out within the
one-boson-exchange model in Ref.~\cite{Li:2012ss}. We look forward to 
an experimental resolution of this issue.

This bleak picture will change dramatically if the $QQ\bar u \bar d$\, TQ is {\em not}
a simple molecular state, but instead a connected color network, consisting of a
$QQ$ diquark coupled to a $\hbox{\bf 3}_c^*$, with a matching 
$\bar u \bar d$ diquark coupled to a {\bf 3}$_c$. The $QQ$ diquark will
then be bound by the large Coulombic interaction ${\cal O}(\alpha_s^2 M_Q)$, 
as discussed below.
Finding such a novel color network
would be particularly exciting \cite{Gelman},\cite{Karliner:2006hf}.  With some poetic
license, we can compare it to  the discovery of the $C_{60}$ buckyballs.

\noindent
The Coulombic binding of the $ QQ $ diquark,
\beq
E_b(QQ)=
{1\over2}\cdot{1\over2}\cdot{1\over 2}{\alpha_s^2} M_Q = {1\over8} \alpha_s^2 M_Q
\label{CoulombicBinding}
\eeq
is half as strong as that of a color-singlet $\bar Q Q$. Another
${1\over2}$ factor reflects the need to use the reduced mass, and the
${1\over2}\alpha_s^2 M_Q$  is the standard hydrogen binding formula. Thus
the heavy  diquark $QQ$ becomes infinitely deeply bound as \hbox{$M_Q \to
\infty$}. For $m_c \sim 1.6$ GeV and using 1/2 for the relevant $\alpha_s$, the
Coulombic binding $\sim 50$ MeV is  moderate.  However, the  $1\over2$ ratio of
QCD Coulomb  interaction in the diquark, versus its strength in the $\bar
Q Q $ color singlet system likely extends to {\em all} QCD interactions,
including the confining stringy potential~\cite{Nussinov:1999sx}.

If indeed the $cc\bar u\bar d$ TQ is mostly described by the new color network,
then its decay into $D D^*$ will be suppressed by the required rearrangement of
the color.

In view of the above discussion, we now focus on the question whether the
lowest state of the new color network is below or near the $D D^*$ threshold.

There are interesting parallels between the 
$QQ\bar q\bar q$ \,TQ-s and doubly-heavy baryons $QQq$. In both types of systems
there is a light color triplet -- a quark or an anti-diquark -- 
bound to a heavy diquark. Because of this similarity, 
{\em experimental observation of doubly-heavy baryons is very important not just
in its own right, but as source of extremely valuable information for
deducing the properties of the more exotic 
$QQ\bar q\bar q$ tetraquarks.} Such deduction can carried out just as it was done for
$b$-baryons.

In the last few years it became possible to accurately predict at the level 
of 2-3 MeV the masses of heavy baryons containing the $b$-quark:
$\Sigma_b(bqq)$, $\Xi_b(bsq)$ and $\Omega_b(bss)$
\cite{KLSigmab,Karliner:2007jp,Karliner:2008sv}.
These predictions used as input the masses of the $B$, $B_s$, $D$ and $D_s$ mesons, together with the masses of the corresponding $c$-baryons
$\Sigma_c(cqq)$, $\Xi_c(csq)$ and $\Omega_c(css)$.

An analogous approach to the masses of the $QQ\bar q \bar q$, suggested the relation
\beq
m(c c\bar u\bar d) = m(\Xi_{ccu}) + m(\Lambda_c) - m(D^0) - {1\over4} [m(D^*)-m(D)]
\label{ccud}
\eeq
designed to optimally match interactions on both sides~\cite{Gelman}.
To date, only the SELEX experiment at Fermilab reported doubly charmed  
$ccd$ and $ccu$ baryons with  mass
 $\sim 3520$ MeV \cite{Mattson:2002vu} -- a result  not confirmed by other
experiments \cite{Aubert:2006qw,Chistov:2006zj}.  Substituting 
this mass in eq.~(\ref{ccud}) as a placeholder in a ``proof of concept" estimate
yields $M(cc\bar u\bar d)$ $\sim 3900$ MeV, which is just 30 MeV 
above the $D^* D$ threshold. 
One should keep in mind that, as implied by eq.~(\ref{CoulombicBinding}),
the binding in $(bb\bar q \bar q)$ is expected to be significantly
stronger than in $(c c\bar q\bar q)$.
In any case, reliable experimental information on $QQq$ baryon masses
is clearly essential for settling the issue.

\section*{Note added}
While this work was being finalized, we became aware of several papers dealing
with closely-related subjects, using different approaches
\cite{Wang:2013cya},\cite{Guo:2013sya},\cite{Chen:2013wca},\cite{Faccini:2013lda}.

\end{document}